\documentclass[12pt]{article}

\usepackage{fullpage,latexsym,amsmath,amsfonts,amssymb,graphicx,tikz}

\newcommand{\absolute}[1]{\left | #1 \right |}
\newcommand{\hilbert}{\mathcal{H}}
\newcommand{\ket}[1]{\left | #1 \right \rangle}
\newcommand{\bra}[1]{\left \langle #1 \right |}
\newcommand{\proj}[1]{\ket{#1} \!\! \bra{#1}}

\newcommand{\amp}[2]{\left \langle #1 | #2 \right \rangle}
\newcommand{\oper}[1]{\boldsymbol{#1}}

\newcommand{\parttr}[1]{\mbox{Tr}_{\subtext{(#1)}}}
\newcommand{\supp}{\mbox{supp}\,}

\newcommand{\sys}[1]{^{\mbox{\tiny (#1)}}}
\newcommand{\subtext}[1]{\mbox{\tiny #1}}
\newcommand{\onehalf}{\mbox{$\frac{1}{2}$}}

\newcommand{\mket}[1]{\left | #1 \right ) }
\newcommand{\mbra}[1]{\left ( #1 \right | }
\newcommand{\mamp}[2]{\left ( #1 \left | #2 \right. \right ) }

\newcommand{\moperation}{\mathcal{E}}
\newcommand{\reduction}[1]{\mbox{\textsf{R}}_{\mbox{\tiny (#1)}}}

\newcommand{\modalspace}{\mbox{$\mathcal{V}$}}
\newcommand{\dual}[1]{{#1}^{\ast}}

\newcommand{\scalarfield}{\mbox{$\mathcal{F}$}}

\newcommand{\vspan}[1]{\left \langle #1 \right \rangle}

\newcommand{\annihilator}[1]{{#1}^{\circ}}

\title{A no-broadcasting theorem for modal quantum theory}
\author{Phillip Diamond, Benjamin Schumacher and Michael D. Westmoreland}

\begin{document}

\maketitle

\begin{abstract}
	The quantum no-broadcasting theorem has an analogue in modal
	quantum theory (MQT), a toy model based on finite fields.  The failure of
	broadcasting in MQT is related to the failure of distributivity of the lattice of
	subspaces of the state space.
\end{abstract}

\section{Introduction}

Modal quantum theory (MQT) \cite{MQT2012,MQT-almostQT} is a toy model of quantum theory.
In MQT, pure states are non-zero vectors $\mket{\psi}$ in 
a finite-dimensional vector space $\modalspace$ over a finite 
field $\scalarfield$.  Evolution of the state over a given time interval is 
described by an invertible operator $\oper{T}$ on $\modalspace$.
A simple measurement is a basis set
$\{ \mket{k} \}$ for $\modalspace$, with each basis vector
identified with a measurement result.  If we write the system state
$\mket{\psi}$ as
\begin{equation}
	\mket{\psi} = \sum_{k} \psi_{k} \mket{k} ,
\end{equation}
then the result $k$ is {\em possible} if and only if $\psi_{k} \neq 0$.
As we see, MQT does not support a full concept of probability, but only
the ``modal'' distinction between possible and impossible 
measurement results.

An equivalent description of measurement makes use of the dual space
$\dual{\modalspace}$.  The basis $\{ \mket{k} \}$ of $\modalspace$ is
associated with a dual basis $\{ \mbra{k} \}$ of $\dual{\modalspace}$ such
that $\mamp{k}{\psi} = \psi_{k}$.  Thus, result $k$ is possible if and only if
$\mamp{k}{\psi} \neq 0$.

Despite the simplicity of MQT, many important features
of actual quantum theory are preserved in the toy model.  Superposition, 
interference, complementarity, entanglement and teleportation
all have analogues in MQT.  The theory also supports versions of a number
of important ``no-go'' theorems of quantum theory.  For instance,
there is a no-cloning theorem in MQT, and the structure of correlations
between entangled systems can be shown to exclude local
hidden variable models \cite{MQT2012}.

In this paper, we will consider the no-broadcasting theorem of
actual quantum theory \cite{nobroadcasting1996}, which is a mixed-state generalization of the
no-cloning theorem.  Our question is whether there is an analogous
result that holds in MQT.  We will find that there is.  Our analysis
raises some unique issues along the way and provides 
new insight into the meaning of the no-broadcasting theorem in
actual quantum theory.

\section{From cloning to broadcasting}

A cloning machine would be a device that takes as input a
quantum system in an unknown state $\ket{\psi}$ and yields
as output two systems in the input state:  $\ket{\psi,\psi}$.
The no-cloning theorem tells us that such a machine
is incompatible with quantum theory.  The first version of the
theorem, due to Wootters and Zurek \cite{wznocloning}, goes like this.  We
suppose that the initial state of the machine is $\ket{M_{0}}$,
and we further suppose that the machine properly clones
distinct input states $\ket{\psi_{a}}$ and $\ket{\psi_{b}}$:
\begin{equation}
	\ket{\psi_{a}, M_{0}} \longrightarrow \ket{\psi_{a},\psi_{a}, M_{a}}
	\quad \mbox{and} \quad
	\ket{\psi_{b}, M_{0}} \longrightarrow \ket{\psi_{b},\psi_{b}, M_{b}}
\end{equation}
where $\ket{M_{a}}$ and $\ket{M_{b}}$ are final machine states.
Now we consider a third state $\ket{\psi_{c}} =  \alpha \ket{\psi_{a}} + \beta \ket{\psi_{b}}$,
a superposition of $\ket{\psi_{a}}$ and $\ket{\psi_{b}}$.  Since the overall
system evolution is linear, it must be that
\begin{equation}
	\ket{\psi_{c},M_{0}} \longrightarrow \alpha \ket{\psi_{a},\psi_{a}, M_{a}}
		+ \beta \ket{\psi_{b},\psi_{b}, M_{b}},
\end{equation}
which is {\em not} the desired state $\ket{\psi_{c}, \psi_{c}, M_{c}}$.
A cloning machine which works for $\ket{\psi_{a}}$ and $\ket{\psi_{b}}$
will fail for the superposition $\ket{\psi_{c}}$.

The second version \cite{ddnocloning} of the theorem considers two inputs 
$\ket{\psi_{a}}$ and $\ket{\psi_{b}}$ that are distinct but
non-orthogonal, so that
$0 < \absolute{\amp{\psi_{a}}{\psi_{b}}} < 1$.  The overall
evolution of the system is unitary, so that 
the inner product of output states must equal that of the input
states.  But
\begin{equation}
	\absolute{ \amp{\psi_{a},\psi_{a}, M_{a}}{\psi_{b},\psi_{b}, M_{b}} }
	= \absolute{\amp{\psi_{a}}{\psi_{b}}}^2 \absolute{\amp{M_{a}}{M_{b}}}
	< \absolute{\amp{\psi_{a},M_{0}}{\psi_{b},M_{0}}} 
	=\absolute{\amp{\psi_{a}}{\psi_{b}}} .
\end{equation}
Thus, the machine cannot clone both $\ket{\psi_{a}}$ and $\ket{\psi_{b}}$.

The first version of the no-cloning theorem states tells us that a 
linearly dependent set of states $\{ \ket{\psi_{a}}, \ket{\psi_{b}}, \ldots \}$
cannot be cloned by any quantum machine.  The second version tells
us that two distinct non-orthogonal states $\ket{\psi_{a}}$ and $\ket{\psi_{b}}$ 
cannot both be cloned.  MQT supports a no-cloning theorem of the first type 
but not the second.  This is because the time evolution of a system 
in MQT is linear but (since $\modalspace$ has no inner product structure) 
not unitary.

In fact, in MQT any two distinct (non-parallel) pure states $\mket{\psi_{a}}$ 
and $\mket{\psi_{b}}$ can be cloned.  This is because the states are distinguishable.  
We can devise a measurement basis that includes both $\mket{\psi_{a}}$ and 
$\mket{\psi_{b}}$.  The cloning machine performs this measurement and 
determines which input state is present, then prepares output states
$\mket{\psi_{a},\psi_{a}}$ or $\mket{\psi_{b},\psi_{b}}$ accordingly.  (Note that
we have described this process in a heuristic ``measure and prepare'' 
way; of course this can be realized as a linear evolution of an extended
system including the machine state.)

The no-broadcasting theorem \cite{nobroadcasting1996} is a generalization of no-cloning 
to mixed states.  Suppose $\rho$ is a density operator describing a single-system state, 
the initial state of system 1.  At the end of the broadcasting process, we have a joint 
state of the two systems 1 and 2.  That is, $\rho\sys{1} \rightarrow \rho\sys{12}$
such that
\begin{equation}
	\parttr{2} \rho\sys{12} = \parttr{1} \rho\sys{12} = \rho .
\end{equation}
This is distinct from cloning because the two output systems are not independent
copies of the original $\rho$.  They may be correlated or entangled.  However,
the final state of each output system, considered by itself, is a faithful reproduction
of the input $\rho$.

Now we can state the no-broadcasting theorem for ordinary quantum theory:
Suppose $\rho_{a}$ and $\rho_{b}$
are two density operators.  Then a broadcasting process for both states
is possible if and only if the operators commute:  $[ \rho_{a}, \rho_{b} ] = 0$.
The second ``unitary'' form of the no-cloning theorem is a special case of
this, since the projections $\proj{\psi_{a}}$ and $\proj{\psi_{b}}$ of distinct 
states commute if and only if they are equal or orthogonal.  And indeed,
when the proof of the no-broadcasting theorem \cite{nobroadcasting1996} 
is examined, it depends
crucially on the fact that the overall evolution of the system is unitary.

\section{What is broadcasting in MQT?}

In MQT, mixed states are not represented by operators, but by subspaces
of the state space $\modalspace$ \cite{MQT-almostQT}.  Given a collection of possible pure states
$\{ \mket{\psi_{a}}, \mket{\psi_{b}}, \ldots \}$, the resulting mixed state $M$
is simply the closed linear span of the possible states, which we denote
\begin{equation}
	M = \vspan{\{ \mket{\psi_{a}}, \mket{\psi_{b}}, \ldots \}} .
\end{equation}
If a measurement is made of the basis $\{ \mket{k} \}$, the result $k$ is
possible provided it is possible for at least one of the pure states in
the mixture.  Equivalently, the result $k$ is impossible if and only if $\mbra{k}$ 
lies within $\annihilator{M}$, the annihilator subspace of $M$ in the dual
space $\dual{\modalspace}$.

Suppose we have a mixed state $M\sys{12}$ of a composite 
system with subsystems 1 and 2.  
Then we can derive mixed states for the individual subsystems in a way
analogous to the partial trace operation in actual quantum theory.
Briefly, these subsystem states are the minimal subspaces $M\sys{1}$ 
and $M\sys{2}$ such that $M\sys{12} \subseteq M\sys{1} \otimes M\sys{2}$.
Suppose $M\sys{12} = \vspan{\mket{\Psi\sys{12}} }$, the one-dimensional
subspace containing $\mket{\Psi\sys{12}}$. Given a basis $\{ \mket{k\sys{1}} \}$
for system 1, we can always write
\begin{equation}
	\mket{\Psi\sys{12}} = \sum_{k} \mket{k \sys{1}} \otimes \mket{\psi_{k}\sys{2}} ,
\end{equation}
where some of the $\mket{\psi_{k}\sys{2}}$ vectors may be zero.  The subsystem
state of system 2 is
\begin{equation}
	M\sys{2} = \vspan{ \{ \mket{\psi_{k}\sys{2}} \} } .
\end{equation}
In the same way we may define the subsystem state $M\sys{1}$.  If the 
original composite state $M\sys{12}$ is a mixed state, the subsystem states
are the minimal subspaces that contain the subsystem states of every
pure state contained in $M\sys{12}$.  Thus, for a pair of modal qubits, if
\begin{equation}
	\mket{\Psi\sys{12}} = \mket{0\sys{1},0\sys{2}} + \mket{1\sys{1},1\sys{2}} ,
\end{equation}
the subsystem states are
\begin{equation}
	M\sys{1} = \vspan{\mket{0\sys{1}},\mket{1\sys{1}}} = \modalspace\sys{1}
	\quad \mbox{and} \quad
	M\sys{2} = \vspan{\mket{0\sys{2}},\mket{1\sys{2}}} = \modalspace\sys{2} .
\end{equation}
This operation of extracting subsystem states is called ``reduction''.  We
would denote this by $M\sys{1} = \reduction{2} M\sys{12}$, etc.

What would constitute a broadcasting process in MQT?  We suppose a set of
possible input states $\{ A, B, \ldots \}$.  Our process
yields two-system output states $M_{A}\sys{12}$, $M_{B}\sys{12}$, etc., such that
\begin{equation}
	\reduction{2} M_{X}\sys{12} = X\sys{1} 
	\quad \mbox{and} \quad
	\reduction{1} M_{X}\sys{12} = X\sys{2}
\end{equation}
for every input $X \in \{ A, B, \ldots \}$.  
The broadcasting process must be realized by linear evolution
on a system including inputs, outputs and apparatus.  Nevertheless, it 
may be convenient to describe it as a measurement followed by a preparation.

In this way, we can show that in MQT any pair of mixed states (subspaces) can
be broadcast.  Consider two subspaces $A$ and $B$ representing the input
mixed states of our process.  Their intersection is a (possibly null)
subspace denoted $R = A \land B$, the ``meet'' of $A$ and $B$.  Their linear
span is another subspace $S = A \lor B$ called their ``join''.  These subspaces
form a lattice
\begin{equation}
\vcenter{\hbox{
\begin{tikzpicture}[scale=1.0]
	\node (A) at (-0.7,0) {$A$};
	\node (B) at (0.7,0) {$B$};
	\node (S) at (0,1) {$S$};
	\node (R) at (0,-1) {$R$};
	\draw (A) -- (S);
	\draw (B) -- (S);
	\draw (A) -- (R);
	\draw (B) -- (R);
\end{tikzpicture}
}}
\end{equation}
where the solid lines represent subspace inclusion.  (That is, $R \subseteq A \subseteq S$
and $R \subseteq B \subseteq S$.)
We can construct a basis
set for $S$ as follows.  First, we find a basis set for the meet $R$:  $\{ \mket{r_{i}} \}$.
This set can be extended to a basis for $A$ by the inclusion of additional vectors
$\{ \mket{a_{j}} \}$ and to a basis for $B$ the inclusion of vectors $\{ \mket{b_{k}} \}$.
The whole collection $\{ \mket{r_{i}}, \mket{a_{j}}, \mket{b_{k}} \}$ is a basis
for $S$.
We now measure the input system in this basis, and conditioned on
the measurement outcome we prepare our output to be two copies of this basis
state.
\begin{itemize}
	\item  If the input is state $A$, only the $\mket{r_{i}}$ and $\mket{a_{j}}$
		basis states represent possible outcomes.  We produce as output
		a mixture of the possible states $\mket{r_{i}\sys{1}, r_{i}\sys{2}}$ and
		$\mket{a_{j}\sys{1}, a_{j}\sys{2}}$.  The state of each subsystem is
		therefore $A = \vspan{\{ \mket{r_{i}}, \mket{a_{j}} \}}$.
	\item  In a similar way, if the input state is $B$, we produce as output 
		a mixture of $\mket{r_{i}\sys{1}, r_{i}\sys{2}}$ states and
		$\mket{b_{k}\sys{1}, b_{k}\sys{2}}$.  Each subsystem is in state $B$.
\end{itemize}

\section{Subspace lattices}

The subspaces of $\modalspace$ form a lattice under the $\land$ and
$\lor$ operations.  Such lattices are always modular, which means that
$A \subseteq B$ implies $A \lor (C \land B) = (A \lor C) \land B$ for any
subspaces $A$, $B$ and $C$ \cite{gratzer}.  However, lattices of subspaces may not be
distributive.  For example, suppose we have two non-parallel vectors
$\mket{a}, \mket{b}$ in a two-dimensional $\modalspace$, and we define 
three subspaces
\begin{equation} \label{eq-rankoneexample}
	A = \vspan{ \mket{a} } \qquad B = \vspan{ \mket{b} } \qquad C = \vspan{ \mket{a} + \mket{b} }.
\end{equation}
Then $C \lor (A \land B) = C$ but $(C \lor A) \land (C \lor B) = \modalspace$.
The operation $\lor$ does not distribute over $\land$.

Our example is a suggestive one.  The three states $\mket{a}$, 
$\mket{b}$ and $\mket{a}+\mket{b}$ cannot be distinguished by
any measurement in MQT.  Furthermore, the MQT version of the no-cloning
theorem tells us that no process can clone all three states.  This
suggests that the mixed-state generalization of no-cloning 
(no-broadcasting) might be connected to the failure of distributivity 
in a lattice of subspaces.

Let us sharpen this conjecture.  Every non-distributive modular lattice
contains a sublattice with the following ``diamond'' structure \cite{gratzer}:
\begin{equation} \label{eq-diamond}
\vcenter{ \hbox{
\begin{tikzpicture}[scale=1.0]
	\node (A) at (-1,0) {$A$};
	\node (B) at (0,0) {$B$};
	\node (C) at (1,0) {$C$};
	\node (S) at (0,1) {$S$};
	\node (R) at (0,-1) {$R$};
	\draw (A) -- (S);
	\draw (B) -- (S);
	\draw (C) -- (S);
	\draw (A) -- (R);
	\draw (B) -- (R);
	\draw (C) -- (R);
\end{tikzpicture} 
} }
\end{equation}
That is, $A \lor B = B \lor C = A \lor C = S$ and $A \land B = B \land C = C \land A = R$.
Our example from Equation~\ref{eq-rankoneexample} 
has exactly this structure, with $R = \vspan{0}$, the null subspace.
That is, its structure is
\begin{equation}\label{eq-nulldiamond}
\vcenter{ \hbox{
\begin{tikzpicture}[scale=1.0]
	\node (A) at (-1,0) {$A$};
	\node (B) at (0,0) {$B$};
	\node (C) at (1,0) {$C$};
	\node (S) at (0,1) {$S$};
	\node (R) at (0,-1) {$\vspan{0}$};
	\draw (A) -- (S);
	\draw (B) -- (S);
	\draw (C) -- (S);
	\draw (A) -- (R);
	\draw (B) -- (R);
	\draw (C) -- (R);
\end{tikzpicture} 
} }
\end{equation}
Now we formally state a proposition, which in Section~\ref{sec-proof} we will
prove as our central result.
\begin{quote}
	{\bf Proposition.}  Suppose the MQT states $A$, $B$ and $C$ 
	lie in a diamond lattice as in Equation~\ref{eq-nulldiamond} .  
	Then no process can broadcast all three states.
\end{quote}
On the other hand, as we will see, if $A$, $B$ and $C$ lie in the lattice
of Equation~\ref{eq-diamond} with $R \neq \vspan{0}$, a broadcasting
process is {\em always} possible.

\section{P-distinguishability}

Before we begin our main proof, we must first introduce generalized measurements
in MQT.  Just as MQT states generalize to mixed state subspaces of $\modalspace$, 
a generalized measurement
is a collection of ``effect'' subspaces $\{ E_{1}, E_{2}, \ldots \}$ of the dual space
$\dual{\modalspace}$such that
\begin{equation}
	\bigvee_{k} E_{k} = \dual{\modalspace} .
\end{equation}
Given a mixed state $M$ in $\modalspace$, the measurement result $k$ is impossible if
$M \subseteq \annihilator{E_{k}}$
and possible otherwise.  Equivalently,
$E_{k}$ is possible if and only if there exist $\mket{\psi} \in M$ and
$\mbra{e} \in E_{k}$ such that $\mamp{e}{\psi} \neq 0$.  (With a slight
abuse of notation, we can denote the existence of such $\mket{\psi}$
and $\mbra{e}$ by writing $E_{k}(M) \neq 0$.)  Any measurement 
procedure, including those involving complex time evolution and
interactions with an ancilla system, can be represented in this way.

Suppose we have a finite set $\{ A_{1}, A_{2}, \ldots , A_{n} \}$ of mixed
states of an MQT system.  These states are distinguishable
if there exists a general measurement $\{ E_{1}, E_{2}, \ldots , E_{n} \}$
such that $E_{j} ( A_{k} ) \neq 0$ if and only if $j = k$.  That is, if one of the
states $A_{k}$ is presented to the measurement, the result would certainly
be $k$.  Here we are interested in a weaker but closely related condition
that we will call {\em p-distinguishability}.  The set $\{ A_{1}, A_{2}, \ldots , A_{n} \}$
is p-distinguishable if there is a measurement that {\em possibly} identifies which
state is present but also might yield a null result.  That is, there exists a
general measurement $\{ E_{1}, E_{2}, \ldots , E_{n}, E_{0} \}$ such that
\begin{itemize}
	\item for all $k = 1, \ldots, n$, $E_{k} (A_{k}) \neq 0$; and
	\item for all $j,k = 1, \ldots, n$ with $j \neq k$, $E_{j} (A_{k}) = 0$.
\end{itemize}
The new feature is the inclusion of the null effect $E_{0}$.
If the state $A_{k}$ is presented to the measurement, the result
must either be $k$ (which is possible) or perhaps the result 0.  
In the former case, we know that the input state is $A_{k}$, but 
in the latter case, we can draw no conclusion.

Let us express p-distinguishability in a different way.  If $j \neq k$, then
for all $\mbra{e} \in E_{j}$ and $\mket{\psi} \in A_{k}$, we must have
$\mamp{e}{\psi} = 0$.  That is, $E_{j} \subseteq \annihilator{A_{k}}$
and $A_{k} \subseteq \annihilator{E_{j}}$.  On the other hand, for every $k$ 
there must exist $\mbra{e} \in E_{k}$ and $\mket{\psi} \in A_{k}$
such that $\mamp{e}{\psi} \neq 0$, so that the result $k$ is possible.
Thus $E_{k} \not\subseteq \annihilator{A_{k}}$ and 
$A_{k} \not\subseteq \annihilator{E_{k}}$.

Consider three subspaces $A$, $B$ and $C$ in the diamond 
lattice of Equation~\ref{eq-diamond}.
These three mixed states cannot be p-distinguishable.  
To see this, consider some candidate measurement
$\{ E_{A}, E_{B}, E_{C}, E_{0} \}$ intended to p-distinguish the three
states.  Then it must be that $E_{C} ( C ) \neq 0$, or in other words
that there exist $\mbra{e} \in E_{C}$ and $\mket{c} \in C$ so
that $\mamp{e}{c} \neq 0$.  Since the state $\mket{c} \in S = A \lor B$,
we can write $\mket{c} = \mket{a} + \mket{b}$ for states in $A$ and $B$.
It follows that $\mamp{e}{c} = \mamp{e}{a} + \mamp{e}{b}$, which
contradicts either the condition $E_{C}(A) = 0$ or $E_{C}(B) = 0$.  
The three states $A$, $B$ and $C$ are thus not p-distinguishable.

We have cast our argument so far in terms of generalized MQT measurements,
and these include processes that involve ancilla systems, interactions, and
so on.  But to emphasize this fact, let us imagine that the states $A$, $B$
and $C$ are first acted upon by some apparatus, yielding processed
states $M_{A}$, $M_{B}$ and $M_{C}$.  These states may not themselves
form a diamond lattice, but they nevertheless remain not p-distinguishable.
This is because any generalized evolution in MQT respects mixtures \cite{MQT-almostQT}.
In other words, an evolution map $\moperation$ that maps input subspaces
to output subspaces must satisfy
\begin{equation}
	\moperation (A \lor B) = \moperation(A) \lor \moperation(B)
\end{equation}
for all subspaces $A$ and $B$.  Since $C \subseteq A \lor B$ in our
diamond diagram, it follows that $M_{C} \subseteq M_{A} \lor M_{B}$.
Hence there cannot exist any measurement on the outputs such
that $E_{C}(M_{A}) = 0$ and $E_{C}(M_{B}) = 0$ but 
$E_{C}(M_{C}) \neq 0$.

\section{Proof of the no-broadcasting theorem for MQT} \label{sec-proof}

To establish our MQT no-broadcasting theorem, therefore, we only need
to demonstrate that broadcast states $M_{A}$, $M_{B}$ and $M_{C}$
produced by some hypothetical machine
must be p-distinguishable.  Let us therefore begin with subspaces
$A$, $B$ and $C$ as in Equation~\ref{eq-diamond}.  We note that 
this diamond structure means that all three of these subspaces must
have the same dimension $d$.  We construct some linearly independent
sets as follows:
\begin{itemize}
	\item  The set $\{ \mket{r_{i}} \}$ is a basis for the space $R$, 
		which has dimension $d_{R}$.  This set is empty if the common
		intersection of $A$, $B$ and $C$ is null, as in Equation~\ref{eq-nulldiamond}.
	\item  The $R$-basis is extended to a basis for $C$ by the addition
		of $d-d_{R}$ vectors $\{ \mket{c_{k}} \}$ that are in $C$ but not in $R$.
		Since $C \subseteq A \lor B$, we can write $\mket{c_{k}} = \mket{a_{k}}
		+ \mket{b_{k}}$ for vectors in $A$ and $B$.  The sets $\{ \mket{a_{k}} \}$
		and $\{ \mket{b_{k}} \}$ must also be linearly independent, and form 
		extensions of the common $R$-basis to basis sets for $A$ and $B$
		respectively.
\end{itemize}
The combined set $\{ \mket{r_{i}}, \mket{a_{k}}, \mket{b_{k}} \}$, for instance, is a 
basis for $S$.

Any vector in $A$ is a linear combination of $\mket{r_{i}}$ and $\mket{a_{k}}$ vectors.
It follows that any vector in the tensor product space $A \otimes A$ is a linear combination of 
$\mket{r_{i}, r_{j}}$, $\mket{a_{k}, r_{j}}$, $\mket{r_{i}, a_{l}}$ and $\mket{a_{k},a_{l}}$ 
vectors.  (We can also make corresponding statements about vectors in $B$, $B \otimes B$,
$C$ and $C \otimes C$.)
The broadcast state $M_{A}$ is a two-system state that reduces to
$A$ on each subsystem.  This means that $M_{A}$ is a subspace of
of $A \otimes A$.  In a similar way, $M_{B} \subseteq B \otimes B$
and $M_{C} \subseteq C \otimes C$.

Now we are ready to propose our p-distinguishing measurement for
$M_{A}$, $M_{B}$ and $M_{C}$.  We define
\begin{equation}
	E_{C} = \annihilator{\left ( (A \otimes A) \lor (B \otimes B) \right )} .
\end{equation}
Similar definitions apply to $E_{A}$ and $E_{B}$, and we 
supplement these by an ``error'' effect $E_{0} = \dual{\modalspace}$.
As defined, the dual subspace $E_{C}$ definitely satisfies the condition that
$E_{C} (M_{A}) = E_{C}(M_{B}) = 0$, because it annihilates both 
$A \otimes A$ and $B \otimes B$.  But can we also guarantee that
$E_{C}(M_{C}) \neq 0$?

First, suppose that the intersection $R = \vspan{0}$.  Then any vector $\mket{\psi}$ in
$M_{C}$ must have the form
\begin{eqnarray}
	\mket{\psi} & = & \sum_{kl} \psi_{kl} \mket{c_{k},c_{l}}  \nonumber \\
	 & = & 
	 \sum_{kl} \psi_{kl} \mket{a_{k},a_{l}} +  \sum_{kl} \psi_{kl} \mket{a_{k},b_{l}} 
	 +  \sum_{kl} \psi_{kl} \mket{b_{k},a_{l}} + \sum_{kl} \psi_{kl} \mket{b_{k},b_{l}} .
\end{eqnarray}
Note the $\mket{a_{k},b_{l}}$ and $\mket{b_{k},a_{l}}$ cross terms.  Such
terms are not present in any vector in $(A \otimes A) \lor (B \otimes B)$.
Thus, $\mket{\psi}$ is not annihilated by $E_{C}$.  It follows (from this and
the parallel arguments for $E_{A}$ and $E_{B}$) that the subspaces
$M_{A}$, $M_{B}$ and $M_{C}$ are p-distinguishable.  Therefore, they 
cannot arise by any dynamical process from the input states $A$, $B$ and $C$.
If the three states form a diamond lattice with null intersection, they cannot
be broadcast.

The situation is very different if $R \neq \vspan{0}$.  Now broadcasting
is simple.  The apparatus makes a random choice to transfer the input state
to one or the other output systems, with the other output chosen to be in state $R$.
That is, our broadcast states are thus
\begin{eqnarray}
	M_{A} & = & (A \otimes R) \lor (R \otimes A) \nonumber \\
	M_{B} & = & (B \otimes R) \lor (R \otimes B) \\
	M_{C} & = & (C \otimes R) \lor (R \otimes C) . \nonumber
\end{eqnarray}
These are not p-distinguishable, but each output system is a duplicate of
the original input.

We should note that an analogous procedure
does not yield a broadcasting operation in actual quantum theory.  Suppose
we have two different states $\rho_{a}$ and $\rho_{r}$.  The first is a possible
input to the broadcasting machine, which is randomly transferred to one of the
two outputs.  The second is the standard state prepared for the other output.
Each output channel ends up in a mixed state
\begin{equation}
	\rho =  \onehalf \rho_{a} + \onehalf \rho_{r},
\end{equation}
which is never the same as the input $\rho_{a}$.  The procedure does not
yield broadcasting.  In MQT, on the other hand, the mixture of states 
$A$ and $R$ is exactly the same as $A$, since the mixture has no ``weighting''.

If the MQT mixed states $A$, $B$ and $C$ form a diamond lattice as in
Equation~\ref{eq-diamond}, even if $R \neq \vspan{0}$, we can consider
subspace states $A_{e} = \vspan{\{ \mket{a_{k}} \}}$, etc., which 
form a diamond lattice with null intersection, as in Equation ~\ref{eq-nulldiamond}.  
These states cannot be broadcast.

\section{Remarks}

In modal quantum theory, the no-broadcasting theorem is not related to a 
failure of commutativity of density operators, which do not exist in MQT.
Instead, it springs from a failure of distributivity in a lattice.  
In actual quantum theory, there is 
a connection among these.  We can associate with any density operator
$\rho$ its ``support'' subspace $\supp \rho$, which is that subspace
spanned by eigenvectors of $\rho$ with non-zero eigenvalues.  Equivalently, 
if we have a mixture of states $\{ \ket{\psi_{n}} \}$ that gives rise to $\rho$,
then $\supp \rho = \vspan{ \{ \ket{\psi_{n}} \} }$.  Now suppose that we
have three states $\rho_{a}$, $\rho_{b}$ and $\rho_{c}$ whose support
subspaces form a diamond lattice:
\begin{equation} \label{eq-fulldiamond}
\vcenter{ \hbox{
\begin{tikzpicture}[scale=1.0]
	\node (A) at (-2,0) {$\supp \rho_{a}$};
	\node (B) at (0,0) {$\supp \rho_{b}$};
	\node (C) at (2,0) {$\supp \rho_{c}$};
	\node (S) at (0,1) {$S$};
	\node (R) at (0,-1) {$R$};
	\draw (A) -- (S);
	\draw (B) -- (S);
	\draw (C) -- (S);
	\draw (A) -- (R);
	\draw (B) -- (R);
	\draw (C) -- (R);
\end{tikzpicture} 
} }
\end{equation}
(Here the subspace $R$ need not be null.)
It follows that at least two of the operators $\rho_{a}$, $\rho_{b}$ and 
$\rho_{c}$ do not commute.  Therefore, we know that the three states
cannot be broadcast.

It is worth recalling that the failure of distributivity in lattices was 
used by Birkhoff and von Neumann  \cite{birkhoffvonneumann1936}
as the basis for non-Boolean 
quantum logic.  In their analysis, ``propositions''
about a quantum system are identified with subspaces of its Hilbert space
$\hilbert$, with meet and join operators identified as logical AND and OR.
Disjunctive propositions (e.g., ``the particle
passed through slit \#1 OR slit \#2'') are true for superposition states, even
though neither of the two component propositions holds.  This yields a 
non-distributive logical structure for quantum propositions.  
Our MQT version of the no-broadcasting theorem, therefore, 
expresses a very fundamental aspect of quantum theory.

In a separate paper \cite{MQTnodeleting}, we will 
consider several other no-go theorems and their possible MQT analogues.  
Some, like the no-deleting theorem, are inherited in MQT; 
others, like the no-hiding theorem, are actually false in the toy model.

PD and BWS gratefully acknowledge the support of the Kenyon Summer
Science Program in the summer of 2022.

\bibliography{Modalquantum}

\begin{thebibliography}{1}

\bibitem{MQT2012}
Benjamin Schumacher and Michael~D. Westmoreland.
\newblock Modal quantum theory.
\newblock {\em Foundations of Physics}, 42:918--925, 2012.

\bibitem{MQT-almostQT}
Benjamin Schumacher and Michael~D. Westmoreland.
\newblock Almost quantum theory.
\newblock In Giulio Chirabella and Robert Spekkens, editors, {\em Quantum
  Theory: Informational Foundations and Foils}, pages 45--81, 2015.

\bibitem{nobroadcasting1996}
Howard Barnum, Carlton~M. Caves, Christopher~A. Fuchs, Richard Jozsa, and
  Benjamin Schumacher.
\newblock Noncommuting mixed states cannot be broadcast.
\newblock {\em Physical Review Letters}, 76:2818, 1996.

\bibitem{wznocloning}
William Wootters and Wojciech Zurek.
\newblock A single quantum cannot be cloned.
\newblock {\em Nature}, 299:802--803, 1982.

\bibitem{ddnocloning}
Dennis Dieks.
\newblock Communication by epr devices.
\newblock {\em Physics Letters A}, 92(6):271--272, 1982.

\bibitem{gratzer}
George Gratzer.
\newblock {\em Lattice Theory: Foundation}.
\newblock Birkhauser, 2011.

\bibitem{birkhoffvonneumann1936}
Garrett Birkhoff and John~Von Neumann.
\newblock The logic of quantum mechanics.
\newblock {\em Annals of Mathematics}, 37(4):823--843, 1936.

\bibitem{MQTnodeleting}
Benjamin Schumacher and Phillip Diamond.
\newblock Cloning, deleting and hiding in modal quantum theory.
\newblock In preparation.

\end{thebibliography}
\bibliographystyle{unsrt}

\end{document}